\documentclass[aps,reprint]{revtex4-1}
\usepackage{graphicx}
\usepackage{amsmath}
\usepackage[svgnames]{xcolor}
\usepackage{gensymb}
\usepackage{lipsum}
\usepackage[version=3]{mhchem}
\usepackage[T1]{fontenc}

\begin{document}

\author{A.~M.~R.~V.~L.~Monteiro}
\email{A.M.Monteiro@tudelft.nl}
\author{D.~J.~Groenendijk}
\author{N.~Manca}
\author{E.~Mulazimoglu}
\author{S.~Goswami}
\author{Ya.~Blanter}
\author{L.~M.~K.~Vandersypen}
\author{A.~D.~Caviglia}
\email{A.Caviglia@tudelft.nl}
\affiliation{Kavli Institute of Nanoscience, Delft University of Technology,\\ P.O. Box 5046, 2600 GA Delft, Netherlands.}

\title{Side gate tunable Josephson junctions at the \ce{LaAlO3}/\ce{SrTiO3} interface}


\begin{abstract}
Novel physical phenomena arising at the interface of complex oxide heterostructures offer exciting opportunities for the development of future electronic devices.
Using the prototypical \ce{LaAlO3}/\ce{SrTiO3} interface as a model system, we employ a single-step lithographic process to realize gate tunable Josephson junctions through a combination of lateral confinement and local side gating. The action of the side gates is found to be comparable to that of a local back gate, constituting a robust and efficient way to control the properties of the interface at the nanoscale. We demonstrate that the side gates enable reliable tuning of both the normal-state resistance and the critical (Josephson) current of
the constrictions. The conductance and Josephson current show mesoscopic fluctuations as a function of the applied side gate voltage, and the analysis of their amplitude enables the extraction of the phase coherence and thermal lengths. Finally, we realize a superconducting quantum
interference device in which the critical currents of each of the constriction-type Josephson junctions can be controlled independently via the side gates.
\end{abstract}

\maketitle


Complex oxide heterostructures host a diverse set of novel physical phenomena which, in combination with an ever-advancing degree of experimental control, shows their promise for fundamental discovery and technological applications\,\cite{hwang2012,cen2009}. Over the past decade, the creation and control of interface superconductivity in oxide heterostructures has attracted a great deal of attention, with special emphasis on the two-dimensional electron system (2DES) hosted at the interface between the two wide band-gap insulators \ce{LaAlO3} (LAO) and \ce{SrTiO3} (STO)\,\cite{ohtomo2004,reyren2007}.  Superconductivity at the LAO/STO interface occurs in an exotic environment with strong spin-orbit coupling\,\cite{caviglia2010,shalom2010,diez2015} in coexistence with localized magnetic moments\cite{bert2011,li2011} and ferroelastic domains\,\cite{kalisky2013,honig2013}. Moreover, the superfluid density can  be tuned using the electrostatic field-effect\,\cite{bert2012}, allowing for an on-off switch of superconductivity by means of an externally applied gate voltage\,\cite{caviglia2008,bell2009}. Despite substantial experimental efforts\,\cite{bert2012,richter2013,cheng2015} accompanied by a growing body of theoretical works\,\cite{michaeli2012,liu2010,banerjee2013}, the microscopic details of superconductivity in the system are still not completely understood.
Efforts to clarify this question have propelled the realization of devices to perform phase-sensitive measurements, opening the possibility to garner information about the symmetry of the superconducting order parameter of the system\,\cite{bert2012,richter2013}. Josephson coupling has recently been reported in STO-based 2DESs, first in top-gated structures where the weak link is electrostatically defined\,\cite{gallagher2014,goswami2015,bal2015}, and later in constriction-type Josephson junctions (c-JJ)\,\cite{goswami2016}. Quantum interference was observed through the integration of two such weak links in a superconducting loop, forming a superconducting quantum interference device (SQUID)\,\cite{goswami2016}. While the top-gating approach benefits from the ability to independently tune each of the weak links, it is rather complex due to the requirement of multiple aligned lithography steps. Moreover, it is well established that the properties of the 2DES at the LAO/STO interface are extremely sensitive to metals and chemicals adsorption\,\cite{dai2016,xie2011,lesne2014} at the surface. These problems can be overcome by employing the simpler c-JJ approach, which requires only a single lithographic step and no further processing after the LAO growth. However, it remains to be shown whether local tunability can be achieved in such weak links.

\begin{figure}[ht!]
\includegraphics[width=\linewidth]{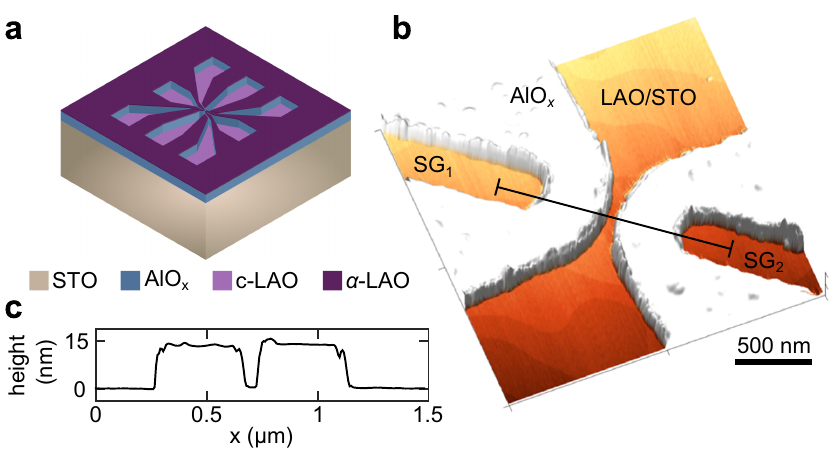}
\caption{\label{fig:fabrication} 
(a) 3D schematic of a side gated constriction. $\alpha$-LAO: amorphous LAO; c-LAO: crystalline LAO.
(b) AFM image of a typical device showing the constriction and two side gates (SG$_1$ and SG$_2$). The 2DES is formed only at the interface between c-LAO and STO.
(c) Height profile along the black line in panel (b), showing a constriction width of approximately $50\,\mathrm{nm}$.}
\end{figure}

In this work, we explore a side gate geometry in order to realize gate-tunable c-JJs at the LAO/STO interface. We demonstrate local electrostatic tunability of these c-JJs while preserving a single lithographic step process by simultaneously defining both the constrictions and the side gate electrodes. 
Similar approaches are often employed in conventional semiconductor based 2DESs to obtain lateral confinement and electrostatically control the effective channel width\,\cite{kristensen2000}. Here, however, we find the electric-field dependence of the STO permittivity to play a crucial role, rendering the action of the side gates comparable to that of an effective ``local back gate''.
For a single junction, phase-coherent transport gives rise to mesoscopic fluctuations of the conductance and of the (Josephson) critical current as a function of side gate voltage. The amplitude of these fluctuations enables us to extract the phase coherence and thermal lengths. Furthermore, we demonstrate the reliability of the side gate electrodes to locally and independently tune the weak links by integrating two side gated c-JJs in a SQUID and controlling the (a)symmetry of its response.


Device fabrication relies on a pre-patterning technique\,\cite{schneider2006,banerjee2012} involving a single lithographic step, which makes use of a template to define the insulating regions on the STO substrate prior to the epitaxial growth of the LAO thin film. Starting from a \ce{TiO2}-terminated STO(001) substrate, we first pattern a resist mask using electron-beam lithography. After development, a thin ($13\,\mathrm{nm}$) \ce{AlO_x} layer is deposited by sputtering and the remaining resist is removed by lift-off in acetone. As a result, the areas of the STO surface protected by the resist during \ce{AlO_x} deposition are cleared, whereas the exposed regions are coated by \ce{AlO_x}. Next, a 12\,u.c.~LAO film is deposited by pulsed laser deposition. In the areas where the STO surface is exposed, the LAO film grows crystalline (c-LAO) and the 2DES forms at the interface. The regions covered by \ce{AlO_x}, where the LAO film grows amorphous ($\alpha$-LAO), remain insulating. The growth process is monitored \textit{in-situ} using reflection high energy electron diffraction (RHEED), displaying a layer-by-layer growth mode. The LAO films were grown at two different temperatures, namely $770^{\circ}$ and $840^{\circ}$. Lower growth temperature results in samples with higher sheet resistance, whereas samples grown at higher temperature exhibit lower sheet resistance and superconductivity.
A more detailed description of the fabrication process can be found in the Supporting Information.
A 3D schematic of a side gated constriction is shown in Figure~\ref{fig:fabrication}a. The \ce{AlO_x} mask delimits the areas where the channel, the bonding pads, and the side gate electrodes are formed. An atomic force microscopy (AFM) image of a constriction and the two side gates is presented in Figure~\ref{fig:fabrication}b. The height profile in Figure~\ref{fig:fabrication}c reveals a channel width ($w$) of approximately $50\,\mathrm{nm}$. We have fabricated different devices varying the distance between the side gates and the conducting channel, and the growth temperature of the LAO thin film. The parameters of the constriction devices presented in this work are summarized in Table~\ref{tab:devices} and all devices show qualitatively similar behavior.
\begin{table}
\centering
\begin{tabular}{c c c c}
			& Dev1	& Dev2	& Dev3 	\\ \hline\hline
 $w$ (nm)	&  50	&  50	& 50	\\
 $L$ (nm) & 500 & 500 & 500 \\
 $d$ (nm)	&  200	& 100	& 100	\\
 $T_{\mathrm{growth}}$ ($^{\circ}$C)	& 770	& 770	& 840 	\\ \hline
\end{tabular}
\caption{Parameters of the side gate devices presented in this work: width ($w$) and length ($L$) of the conducting channel, distance between the channel and the side gates ($d$) and the growth temperature of the LAO thin film ($T_{\mathrm{growth}}$).} \label{tab:devices}
\end{table}


\begin{figure}[ht!]
\includegraphics[width=\linewidth]{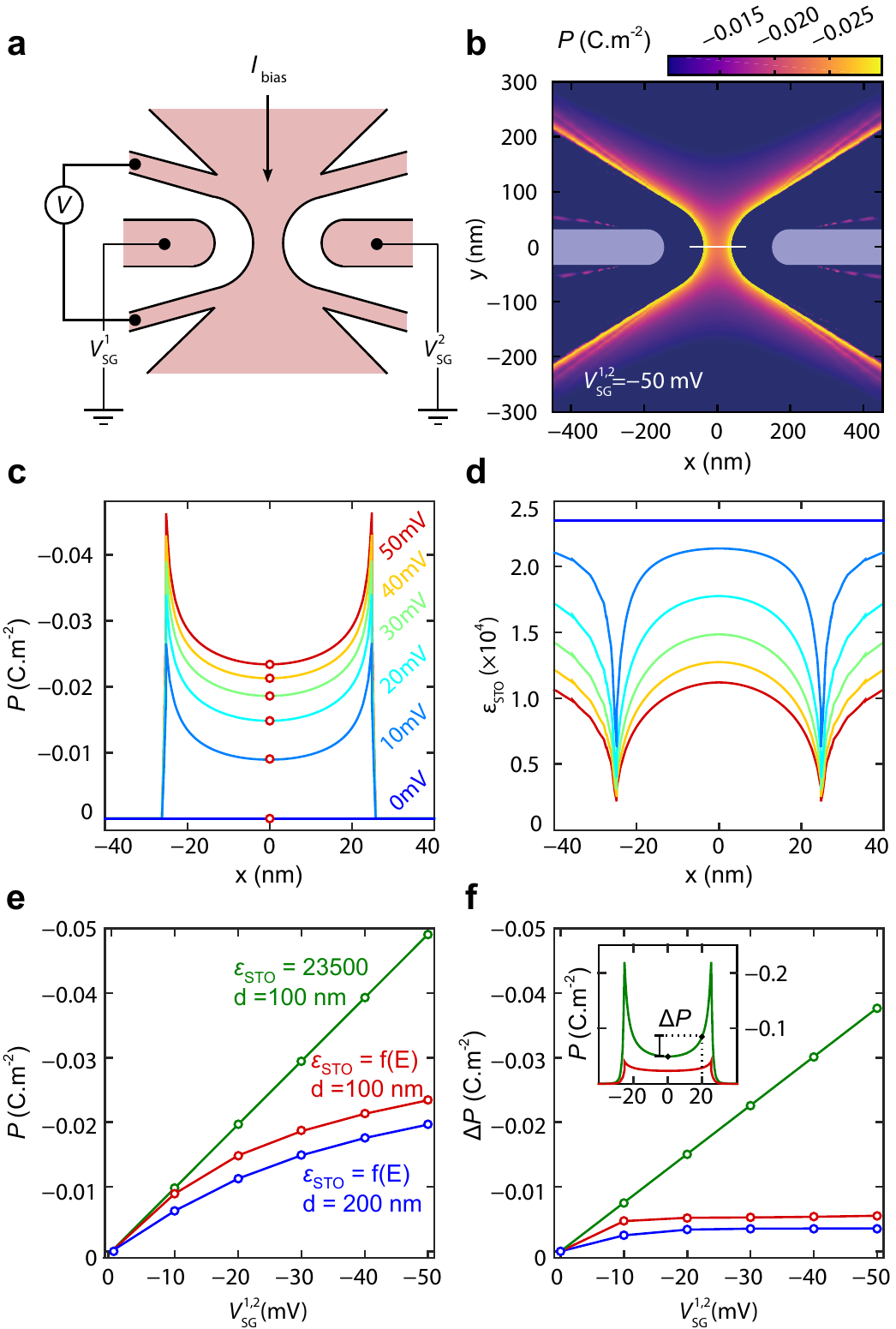}
\caption{\label{fig:simulations} 
(a) Sketch of the device geometry showing the electrical connections for transport measurements. 
(b) Spatial map of the out-of-plane electric polarization ($P$) for $V^{1,2}_{\mathrm{SG}}=-50\,\mathrm{mV}$, obtained by finite-element simulations. 
(c) and (d) Evolution of $P$ and $\varepsilon_{\mathrm{STO}}$, respectively, across the constriction (along the white line in panel b) for different values of $V^{1,2}_{\mathrm{SG}}$. 
(e) Value of $P$ at the center of the constriction ($x=0\,\mathrm{nm}$) as a function of $V^{1,2}_{\mathrm{SG}}$. 
(f) $\Delta P$ as a function of $V^{1,2}_{\mathrm{SG}}$. Color code as in panel e. Inset: electric polarization profiles across the constriction. 
}
\end{figure}

The measurement configuration used is shown in Figure~{\ref{fig:simulations}}a. A constant DC current ($I_\mathrm{bias}$) is injected through the conducting channel and the voltage drop ($V$) is locally measured at the constriction using two probes on the side. Voltages can be applied independently to the two side gates ($V^1_\mathrm{SG}$ and $V^2_\mathrm{SG}$), enabling local modulation of transport across the constriction by field-effect. 
In order to understand how the local side gates modulate transport through the constriction in this geometry, we performed Finite Element Analysis (FEA) in COMSOL\textregistered~(see Supporting Information for modeling specific details). 
Calculations are performed for the geometry sketched in Figure~\ref{fig:simulations}b, using a channel width $w \approx 50\,\mathrm{nm}$ and a distance $d \approx 100\,\mathrm{nm}$ between the side gates and the channel.
An important aspect that has to be addressed is the role of the strong electric-field dependence of the permittivity of the STO substrate, which is not commonly found in other systems.
Its electric-field dependence is modeled as\,\cite{landau1981,ang2004} 
\begin{equation}
\varepsilon_{\mathrm{STO}}(E) = 1+\frac{B}{[1+(E/E_0)^2]^{1/3}}
\end{equation}
with $B = 23,500$ and $E_0 = 82,000\,\mathrm{V/m}$\,\cite{stornaiuolo2014}. The side gate electrodes are modeled as areas of fixed voltage and the conducting channel as a ground plane. This approximation is valid provided the voltage drop across the constriction is negligible when compared to the magnitude of the voltages applied to the side gate electrodes. 

Figure~\ref{fig:simulations}b shows a spatial map of the calculated out-of-plane electric polarization ($P$) in a symmetric gating configuration with $V^{1,2}_\mathrm{SG} = -50\,\mathrm{mV}$. The out-of-plane polarization is directly related to the accumulated charge carrier density at the interface by $\Delta n_\mathrm{2D} = P/e$. Due to the coplanar capacitor arrangement, crowding of electric-field lines occurs at the edges of the constriction. Figure~\ref{fig:simulations}c shows the resulting variation of $P$ across the channel (white line in Figure~\ref{fig:simulations}b) for different values of $V^{1,2}_\mathrm{SG}$. For all the curves, the magnitude of $P$ is maximum at the edges of the channel and minimum at its center. In turn, the enhancement of the local electric-field intensity at the edges of the constriction results in a large decrease of the $\varepsilon_\mathrm{STO}$ as shown in Figure~\ref{fig:simulations}d. This has two consequences that affect the electrostatic gating mechanism. Firstly, it leads to a progressive saturation of the depleted carrier density in the channel, reducing the gating efficiency as $V^{1,2}_\mathrm{SG}$ increases in magnitude. Secondly, it redistributes the electric-field lines towards the centre of the constriction, flattening out the depletion profile.
The first effect is reported in Figure~\ref{fig:simulations}e where  the calculated polarization at the center of the channel ($P_{x=0\,\mathrm{nm}}$) is plotted as a function of $V^{1,2}_\mathrm{SG}$. The electric-field dependence of $\varepsilon_{\mathrm{STO}}$ produces a deviation from the linear behavior one would obtain for $\varepsilon_{\mathrm{STO}} = \mathrm{constant}$, resulting in a reduction of the depleted charge. 
Moreover, within the same side gate voltage range, the magnitude of the induced polarization is smaller for the case $d=200\,\mathrm{nm}$ than for $d=100\,\mathrm{nm}$. This highlights the fact that the charge depletion is strongly dependent on the distance between the side gate electrodes and the channel.
In Figure~\ref{fig:simulations}f we show how the steepness of the polarization profile across the channel is affected by the field dependence of $\varepsilon_{\mathrm{STO}}$. To evaluate this effect we consider the variation of the polarization between $x=0\,\mathrm{nm}$ and $x=20\,\mathrm{nm}$ , $\Delta P = P_{x=20\,\mathrm{nm}} - P_{x=0\,\mathrm{nm}}$.  For the case $\varepsilon_{\mathrm{STO}}=\mathrm{constant}$, the magnitude of $\Delta P$ increases linearly with applied side gate voltage, representing a proportional scaling of the polarization profile. In contrast, for the case $\varepsilon_{\mathrm{STO}}=f(E)$, it rapidly saturates at $\sim 5\,\mathrm{\mu C/cm^{2}}$. This limits the extent to which side gating can reduce the effective width of the channel, since the depletion profile is less steep due to the dielectric response of the STO, as opposed to the stronger sideways depletion that would be obtained if $\varepsilon_{\mathrm{STO}}=\mathrm{constant}$ (red and green curves in the inset of Figure~\ref{fig:simulations}f, respectively).
This is in good agreement with previous reports on side gate electrodes at the LAO/STO interface, which show the electric field lines reaching the gas from below\,\cite{stornaiuolo2014}. Therefore, the side gates are expected to act as an effective ``local back gate'' due to the proximity to the channel and can be used to locally modulate the carrier density at the nanoscale with the application of small voltages.


\begin{figure}[ht!]
\includegraphics[width=\linewidth]{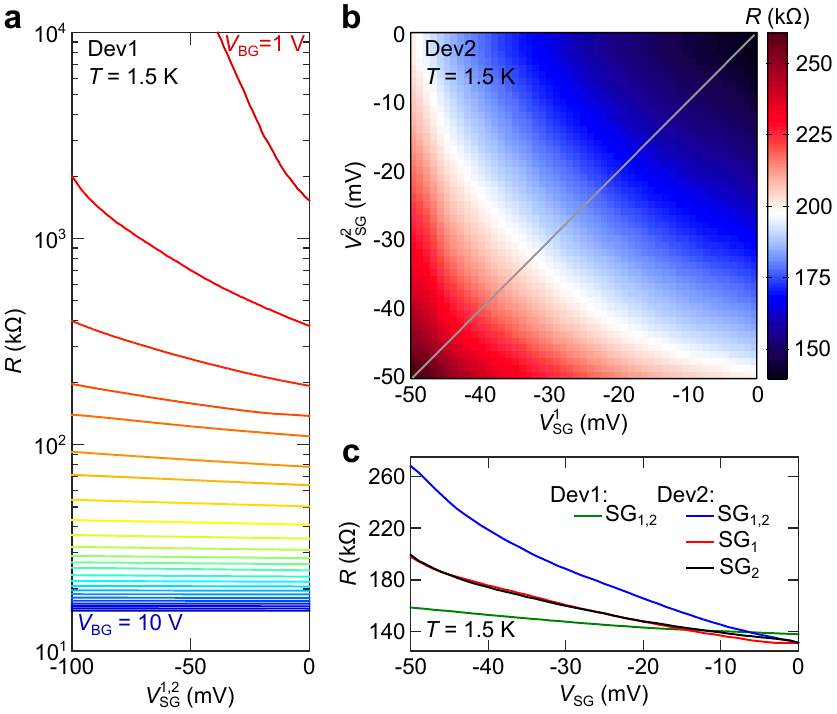}
\caption{\label{fig:R} 
(a) 4-probe resistance ($R$) of as a function of side gate voltage $V^{1,2}_{\mathrm{SG}}$ measured for different $V_{\mathrm{BG}}$.
(b) Map of $R$ as a function of $V^1_{\mathrm{SG}}$ and $V^2_{\mathrm{SG}}$. The  voltage step is $0.2\,\mathrm{mV}$. 
(c) $R$ as a function of side gate voltage. Dev1: $V^1_{\mathrm{SG}}=V^2_{\mathrm{SG}}$ (green). Dev2: $V^1_{\mathrm{SG}}=V^2_{\mathrm{SG}}$ (blue), $V^2_{\mathrm{SG}}=0\,\mathrm{mV}$ (red) and $V^1_{\mathrm{SG}}=0\,\mathrm{mV}$ (black).
}

\end{figure}

We first characterize the constrictions and the action of the two side gates through transport measurements in the normal state, at $1.5\,\mathrm{K}$. Figure~\ref{fig:R} shows electrical measurements of devices Dev1 and Dev2, where the side gates are separated from the constriction by $200\,\mathrm{nm}$ and $100\,\mathrm{nm}$, respectively. In both devices, the overall carrier density can be tuned by the application of a back gate voltage ($V_{\mathrm{BG}}$). The evolution of the 4-probe resistance ($R$) across the constriction in Dev1 as a function of $V^{1,2}_{\mathrm{SG}}$ for different $V_{\mathrm{BG}}$ is shown in Figure~\ref{fig:R}a. At $V_{\mathrm{BG}}=10\,\mathrm{V}$, $R$ remains fairly unchanged within the side gate voltage range considered. This results from the fact that the change in carrier density $\Delta n_{\mathrm{2D}}$ induced by the side gates is a small fraction of the total carrier density accumulated by the back gate. As the back gate voltage is decreased, the effect of the side gates becomes increasingly appreciable and at $V_{\mathrm{BG}}=1\,\mathrm{V}$, the side gates can completely deplete the channel. 

Having established a range of $V_{\mathrm{BG}}$ in which the voltages applied to the side gates induce appreciable changes in the transport through the constriction, we now study the individual action of the side gates. 
Figure~\ref{fig:R}b shows a 2D map of the 4-probe resistance across the constriction of Dev2 as a function of $V^1_{\mathrm{SG}}$ and $V^2_{\mathrm{SG}}$. The action of each side gate on the constriction is identical, evidenced by the symmetry across the diagonal (gray line). This is also reflected in the good overlap between the red and black curves in Figure~\ref{fig:R}c, measured as SG$_1$ and SG$_2$ were individually driven from 0 to $-50\,\mathrm{mV}$, while keeping the other side gate at $0\,\mathrm{mV}$. 
When both SG$_{1}$ and SG$_{2}$ are swept symmetrically, the resistance change is roughly twice as large (blue curve).
This underlines the reliability of the patterning technique, where the action of the side gates is determined by the geometrical design.
When comparing the induced change in resistance as a function of $V^{1,2}_{\mathrm{SG}}$ for Dev1 and Dev2, we observe it to be much smaller for Dev1, where the side gate electrodes are patterned further away from the constriction. This corroborates the expectation that the tunability of the resistance strongly depends on $d$, in good agreement with the simulations from Figure\,\ref{fig:simulations}e and f.


\begin{figure}[ht!]
\includegraphics[width=\linewidth]{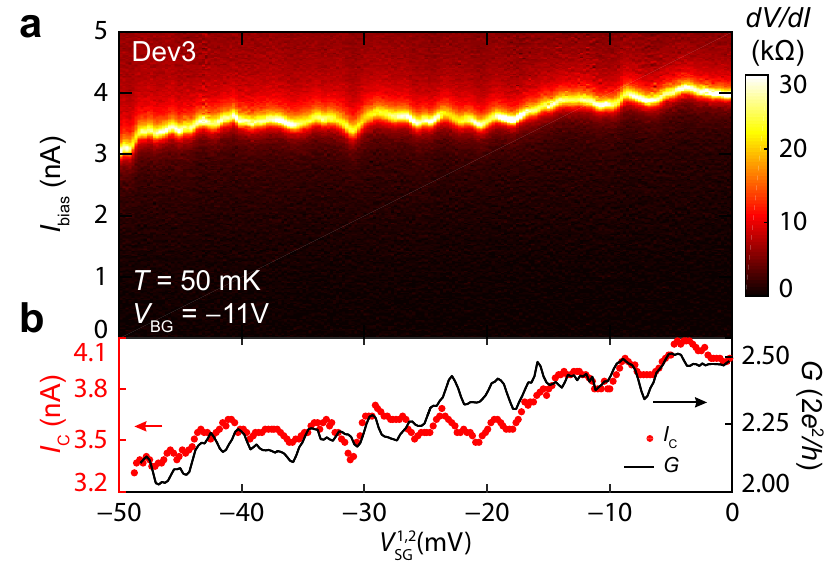}
\caption{\label{fig:SC1} 
(a) Differential resistance ($dV/dI$) plotted as a function of bias current $I_{\mathrm{bias}}$ and side gate voltage $V^{1,2}_{\mathrm{SG}}$, measured at $V_{\mathrm{BG}}=-11\,\mathrm{V}$ and $T=50\,\mathrm{mK}$. 
(b) Fluctuations of the conductance $G$ and the critical current $I_{\mathrm{c}}$ as a function of applied side gate voltage $V^{1,2}_{\mathrm{SG}}$. 
}
\end{figure}

We now turn our attention to the study of the superconducting regime. In previous work it was shown that such constrictions act as a weak link between the two superconducting reservoirs, forming a Josephson junction (c-JJ type)\,\cite{goswami2016}.
We first focus on the study and side gate modulation of transport through a single Josephson junction.
In Figure~\ref{fig:SC1}a, the differential resistance $dV/dI$ is plotted in color scale as a function of bias current $I_{\mathrm{bias}}$ and side gates voltage $V^{1,2}_{\mathrm{SG}}$, i.e., in the symmetric side gating configuration (see Supporting information for the study as a function of the independent side gate voltages). The measurement was performed at $T=50\,\mathrm{mK}$ with a fixed back gate voltage $V_{\mathrm{BG}}=-11\,\mathrm{V}$. 
It can be seen that, on average, the supercurrent range (black region) is reduced when $V^{1,2}_{\mathrm{SG}}$ is driven to larger negative values, due to the consequent decrease of the electron carrier density.
The corresponding values of critical current $I_{\mathrm{c}}$ as a function of $V^{1,2}_{\mathrm{SG}}$ are plotted in Figure~\ref{fig:SC1}b.  In the same graph, the differential conductance $G$ is plotted in units of $2e^2/h$. $I_{\mathrm{c}}$ and $G$ were determined from the differential resistance $dV/dI$, from the position of its maximum and its value at the maximum $I_\mathrm{bias}$, respectively (see Supporting Information).
It can be seen that both $I_{\mathrm{c}}$ and $G$ fluctuate in an aperiodic manner as a function of applied side gate voltage. Measurements over several sweeping cycles (see Supporting Information) indicate that these fluctuations are time-independent and reproducible within the sample. Moreover, we observe that the fluctuation pattern of $I_{\mathrm{c}}$ is similar to that of $G$, indicating a common physical origin. We note the absence of such mesoscopic fluctuations in Dev2 at $1.5\,\mathrm{K}$, due to the low $T_{\mathrm{growth}}$ and consequent high sheet resistance.

The mesoscopic fluctuations of the critical current and conductance -- so-called universal conductance fluctuations (UCF)-- originate from the phase-coherent transport in a system with dimensions comparable to the phase coherence length ($L_\phi$), where only a limited number of inelastic scattering centres are involved\,\cite{lee1987}. 
Previous transport studies\,\cite{stornaiuolo2012} at the LAO/STO interface have yielded a coherence length $L_\phi \sim 110\,\mathrm{nm}$ at $40\,\mathrm{mK}$. From the AFM image we estimate a junction length $L \approx 500\,\mathrm{nm}$ (see Supporting Information), therefore our c-JJs are expected to display mesoscopic fluctuations.

From the theoretical point of view, fluctuations in the critical current of a Josephson junction have been treated in two limits.
For the short junction limit, where the junction length is much shorter than the superconductor coherence length ($L\ll\xi$), Beenakker\,\cite{beenakker1991} has shown that the fluctuations are universal and that their amplitude depends only on the superconducting gap $\Delta$: $\delta I^{\mathrm{rms}}_{\mathrm{c}} \sim e \Delta/h$. One does not expect our devices to fit this regime, since the measured length of the constriction $L \approx 500\,\mathrm{nm}$ is larger than previously reported values\,\cite{reyren2007,reyren2009} $\xi \sim 100\,\mathrm{nm}$ (at optimal doping).
In the long junction limit, the reduction in the amplitude of the critical current fluctuations is caused by classical self-averaging due to inelastic scattering events, i.e., $L_{\phi}<L$. In addition, at finite temperatures thermal averaging needs to be taken into consideration when $k_{\mathrm{B}}T > hD/L^{2}_{\mathrm{\phi}}$, where $D$ is the diffusion constant. The length associated with this dephasing mechanism is the thermal length $L_{\mathrm{T}} = \sqrt{hD/k_{\mathrm{B}}T}$.
Al'tshuler and Spivak\,\cite{altshuler1987} have considered the theoretical description of the oscillations in critical current for the case $L_{\phi}\gg L \gg L_{\mathrm{T}}$. Since the aforementioned dephasing mechanisms should contribute to the reduction of the amplitude of mesoscopic oscillations in the same way, we consider an effective length $L_{\mathrm{eff}} = (L_{\mathrm{T}}^{-2} + L_{\phi}^{-2} )^{-1/2}$.
Then, for the limit $L_{\mathrm{eff}} \gg w, t$, where $t \approx 10\,\mathrm{nm}$ is the thickness of the 2DES\,\cite{reyren2009}, the root mean square of the oscillations in critical current becomes\,\cite{altshuler1987}
\begin{equation}
\delta I^{\mathrm{rms}}_\mathrm{c} = \frac{4ek_{\mathrm{B}}T}{h} \cdot \sqrt{\exp \left( \frac{-2L}{L_{\mathrm{eff}}}\right) \cdot (2\pi)^{5/2}\frac{L}{L_{\mathrm{eff}}}}.
\end{equation}
From the experimentally observed value $\delta I^{\mathrm{rms}}_{\mathrm{c}}=0.09\,\mathrm{nA}$ we extract $L_{\mathrm{eff}} \approx 100\,\mathrm{nm}$, which provides a relation between $L_\mathrm{T}$ and $L_\phi$.

The amplitude of the conductance fluctuations now enables us to extract numerical values for these two length scales.
In the microscopic theory of Lee, Stone, and Fukuyama\,\cite{lee1987}, the root mean square of the conductance oscillations ($\delta G^{\mathrm{rms}}$) was  evaluated analytically only in the asymptotic regimes of $L_{\phi} \ll L_{\mathrm{T}}$ and $L_{\mathrm{T}} \ll L_{\phi}$. At the LAO/STO interface, however, these
two characteristic length scales are comparable\,\cite{rakhmilevitch2010}, namely, $L_{\phi} \sim L_{\mathrm{T}}$. In order to facilitate comparison with the quasi-1D limit ($w< L_{\mathrm{T}}, L_{\phi} < L$), Beenakker and van Houten have proposed an approximate formula to interpolate between the two asymptotic regimes
\begin{equation}
\delta G^{\mathrm{rms}} = \alpha \cdot \left( \frac{e^2}{h} \right) \cdot \left( \frac{L_\phi}{L_x} \right)^{3/2} \cdot \left[ 1 + (\alpha^2/\beta^2)(L_\phi/L_T)^2 \right]^{-1/2}.
\end{equation}
We take $\alpha = \beta = 0.73$, which recover the asymptotic results originally obtained in Ref.\,\cite{lee1987}.
From the data in Figure\,\ref{fig:SC1}b we obtain $\delta G^{\mathrm{rms}}=0.086 e^2/h$ (see Supporting Information), which, together with $L_{\mathrm{eff}}=100\,\mathrm{nm}$,  yields $L_{\phi} \approx 170\,\mathrm{nm}$ and $L_{\mathrm{T}} \approx 120\,\mathrm{nm}$. Hence, we can estimate a diffusion constant $D \approx 0.16\,\mathrm{cm^2/s}$ and a Thouless energy $E_{\mathrm{Th}}= D \hbar / L_{\phi}^2 \approx 0.4\,\mathrm{\mu eV}$. 
As previously mentioned, we expect the studied device to belong to the long junction limit based on the value of $L$ estimated from the AFM image. In this regime, the Thouless energy should be the dominant energy scale, i.e., $E_{\mathrm{Th}} \ll \Delta$. Within the range of side gate voltages considered, $e I_{\mathrm{c}}R \approx 20\,\mathrm{\mu eV}$, which allows us to estimate $\Delta \approx 7$ - $14\,\mathrm{\mu eV}$. Hence, the value of $E_\mathrm{Th}$ determined from the combined analysis of the critical current and conductance fluctuations is in good agreement with the long junction limit.

\begin{figure}[ht!]
\includegraphics[width=\linewidth]{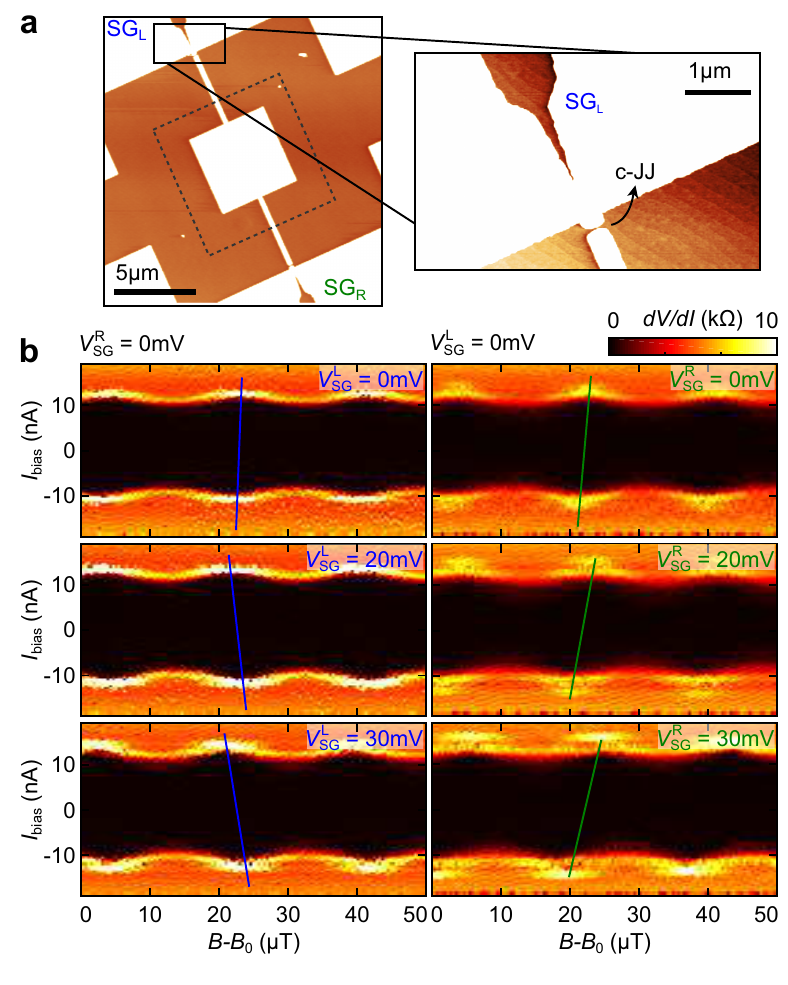}
 \caption{\label{fig:SQUID}
(a) AFM image of the SQUID device which comprises a left (SG$_{\mathrm{L}}$) and right (SG$_{\mathrm{R}}$) side gate electrodes. Inset: c-JJ of the left arm and the the respective side gate electrode.
(b) Tunability of the the SQUID oscillations. Left column: $V_{\mathrm{SG}}^{\mathrm{R}} = 0\,\mathrm{mV}$ and different values of $V_{\mathrm{SG}}^{\mathrm{L}}$. Right column: $V_{\mathrm{SG}}^{\mathrm{L}}= 0\,\mathrm{mV}$ and different values of $V_{\mathrm{SG}}^{\mathrm{R}}$. $B_0$ is an experimentally determined offset in the magnetic field and has an uncertainty greater than one oscillation period.
}
\end{figure}

Finally, we demonstrate the tunability offered by the side gates by integrating two side gated c-JJs in a superconducting loop to create a SQUID. Figure\,\ref{fig:SQUID}a shows an AFM image of the device which comprises a left (SG$_{\mathrm{L}}$) and right (SG$_{\mathrm{R}}$) side gate electrode to allow independent control of each c-JJ. If an external magnetic field is applied perpendicularly to the superconducting loop, the measured critical current oscillates periodically with the changes in phase at the two c-JJ junctions, as seen in Figure\,\ref{fig:SQUID}b. We plot the differential resistance in color scale as a function of current bias ($I_{\mathrm{bias}}$) and applied (out-of-plane) magnetic-field ($B$) for different combinations of side gate voltages. We start by investigating the case when no voltages are applied to the side gate electrodes (top panel), where periodic oscillations of the critical current as a function of magnetic-field are observed. The period of these oscillations is $\Delta B = 19\,\mathrm{\mu T}$, which yields an effective loop area $A_{\mathrm{eff}}= h/2e\Delta B$ of approximately  $8 \times 8\,\mathrm{\mu m^2}$, indicated by the dashed line in Figure\,\ref{fig:SQUID}a. The difference between the estimated effective area and the area of the central insulating region ($5 \times 5\,\mathrm{\mu m^2}$) is expected, and originates from flux-focusing effects due to the fact that the dimensions of the SQUID are smaller than the Pearl length ($\sim 1\,\mathrm{mm}$)\,\cite{goswami2016}. However, a small offset along the $B$-axis can be observed between the oscillations of positive and negative critical current. 
This asymmetry arises due to self flux effects, which are particularly important for SQUIDs with a large kinetic inductance ($L_{\mathrm{k}}$). This is indeed the case for SQUIDs made at the LAO/STO interface\,\cite{goswami2016}, where the low superfluid density results in an exceptionally large kinetic inductance.
The tunability of the superfluid density by electric field-effect therefore provides a direct way to modulate the kinetic inductance through the application of a gate voltage. In our device, the reduced width of the c-JJs causes the kinetic inductance of each arm to be dominated by the weak link. The local modulation of the side gates therefore simultaneously affects the critical current of the weak link and the kinetic inductance of the arm. Thus, the independent tunability of each c-JJ should allow for the control of the asymmetry in the positive and negative critical current oscillations. By keeping $V_{\mathrm{SG}}^{\mathrm{R}}$ fixed at  $0\,\mathrm{mV}$ while driving $V_{\mathrm{SG}}^{\mathrm{L}}$ to positive values (left column), the offset of the positive and negative $I_{\mathrm{c}}(B)$ along the $B$ axis increases, as denoted by the blue lines connecting two maxima of both branches. In turn, when $V_{\mathrm{SG}}^{\mathrm{L}}$ is kept fixed at  $0\,\mathrm{mV}$ and $V_{\mathrm{SG}}^{\mathrm{R}}$ made more positive (right column), the offset occurs in the opposite direction, as expected. This underlines the reliability of the side gate geometry in providing independent modulation of the c-JJs, thus enabling the control of the SQUID asymmetry.

In summary, we have realized nanoscale constrictions at the LAO/STO interface in conjunction with side gate electrodes, which are patterned in the 2DES itself, allowing for a single lithography step process. We have shown that such side gates allow for the reliable local modulation of transport across the constriction by electric field-effect. Finite element simulations show that, due to the electric-field dependence of the permittivity of STO, the action of these side gate electrodes is comparable to that of an effective ``local back gate''. 
Transport measurements in the normal state have corroborated the reliability of the pre-patterning technique used, by demonstrating a symmetric action of both side gates. In the superconducting regime, mesoscopic oscillations of conductance and Josephson supercurrent allow for the estimation of $L_{\phi}$ and $L_{\mathrm{T}}$. Lastly, we integrate two side gated c-JJs in a superconducting loop to realize a SQUID. The subsequent control of the (a)symmetry of the SQUID response via the side gate electrodes underscores the reliability of our single-step technique. We demonstrate efficient local electrostatic control of the c-JJs,
with the additional advantage of not requiring any post processing after the LAO growth.
The results reported in this work open exciting perspectives for the study of quasi-one dimensional superconductivity and for the realization of devices such as superconducting quantum point contacts. 

\section*{Acknowledgment}
The authors thank G.~Steele, H.S.J.~van der Zant and A.~Akhmerov for useful discussions and Tino Kool and Ronald Bode for technical support. 
This work was supported by The Netherlands Organisation for Scientific Research (NWO/OCW) as part of the Frontiers of Nanoscience program and by the Dutch Foundation for Fundamental Research on Matter (FOM).

\bibliography{references}


%
%

\end{document}